# Personalized Recommendation via Integrated Diffusion on User-Item-Tag Tripartite Graphs

**Zi-Ke Zhang, Tao Zhou and Yi-Cheng Zhang**

*Department of Physics, University of Fribourg, Chemin du Musée 3, CH-1700, Fribourg,  Switzerland*

**Abstract**

Personalized recommender systems are confronting great challenges of accuracy, diversification and novelty, especially when the data set is sparse and lacks accessorial information, such as user profiles, item attributes and explicit ratings. Collaborative tags contain rich information about personalized preferences and item contents, and are therefore potential to help in providing better recommendations. In this paper, we propose a recommendation algorithm based on an integrated diffusion on user-item-tag tripartite graphs. We use three benchmark data sets, Del.icio.us, MovieLens and BibSonomy, to evaluate our algorithm. Experimental results demonstrate that the usage of tag information can significantly improve accuracy, diversification and novelty of recommendations

Keywords: Personalized Recommendation; Collaborative Tagging; Diffusion; Tripartite Graphs; Folksonomy

## 1.    Introduction

The last few years have witnessed an explosion of information that the exponential growth of the Internet [1] and World Wide Web [2] confronts us with an information overload: We face too much data and sources to be able to find out those most relevant for us. Indeed, we have to make choices from thousands of movies, millions of books, billions of web pages, and so on. Evaluating all these alternatives by ourselves is not feasible at all. As a consequence, an urgent problem is how to automatically find out the relevant items for us. Internet search engine [3], with the help of keyword-based queries, is an essential tool in getting what we want from the web. However, the search engine does not take into account personalization and returns the same results for people with far different habits. In addition, not all needs or tastes can be easily presented by keywords. Comparatively, recommender system [4], which adopts knowledge discovery techniques to provide personalized recommendations, is now considered to be the most promising way to efficiently filter out the overload information. Thus far, recommender systems have successfully found applications in e-commerce [5], such as book recommendations in Amazon.com [6], movie recommendations in Net°ix.com [7], video recommendations in TiVo.com [8], and so on.

*Correspondence to*: Zi-Ke Zhang. Department of Physics, University of Fribourg, Chemin du Musée 3, CH-1700, Fribourg, Switzerland. Email: zhangzike@gmail.com.

# Zi-Ke Zhang, Tao Zhou and Yi-Cheng Zhang

A recommender system is able to automatically provide personalized recommendations based on the historical record of users' activities. These activities are usually represented by the connections in a user-item bipartite graph [9, 10]. Figure 1 illustrates such a graph consisted of five users and four books, where users can give ratings to those books. So far, collaborative Filtering (CF) is the most successful technique in the design of recommender systems [11], where a user will be recommended items that people with similar tastes and preferences liked in the past. Despite its success, the performance of CF is strongly limited by the sparsity of data resulted from: (i) the huge number of items far beyond user's ability to evaluate even a small fraction of them; (ii) users do not incentively wish to rate the purchased/viewed items [12]. As shown in Figure 1, besides the fundamental user-item relations, some accessorial information can be exploited to improve the algorithmic accuracy [13]. User profiles, usually including age, sex, nationality, job, etc., can be treated as prior known information to filter out possibly irrelevant recommendations [14], however, the applications are mostly forbidden or strongly restricted to respect personal privacy. Attribute-aware method [15] takes into account item attributes, which are defined by domain experts. Yet it is limited to the attribute vocabulary, and, on the other hand, attributes describe global properties of items which are essentially not helpful to generate personalized recommendations. In addition, content-based algorithms can provide very accurate recommendations [16], however, they are only effective if the items contain rich content information that can be automatically extracted out, for example, these methods are suitable for recommending books and articles, but not for videos or pictures. Collaborative tagging systems (CTSes), allowing users to freely assign tags to their collections, provide promising possibility to better address the above issues. CTSes require no specific skills for user participating, thus can overcome the limitation of vocabulary domains and size, widen the semantic relations among items and eventually facilitate the emergence of *folksonomy* [17]. Actually, tags can be treated as abstracted content of items with personalized preferences. In this paper, we propose an integrated diffusion-based recommendation algorithm making use of the ternary relations among users, items and tags. We use three benchmark data sets, *Del.icio.us*, *MovieLens* and *BibSonomy*, to evaluate our algorithm. Experimental results demonstrate that the usage of tag information can significantly improve accuracy, diversification and novelty of recommendations.

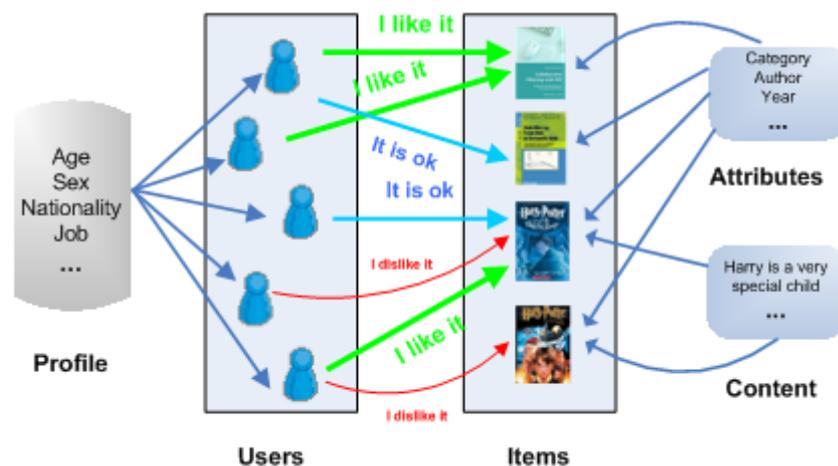

Fig.1. Illustration of a recommender system consisted of five users and four books. The basic information contained by every recommender system is the relations between users and items that can be represented by a bipartite graph. This illustration also exhibits some additional information frequently exploited in the design of recommendation algorithms, including user profiles, item attributes and item content.

Recently, many efforts have been addressed in understanding the structure and evolution of CTSes [18, 19], as well as the usage patterns in folksonomies [20]. A considerable number of algorithms are designed to recommend tags to users, which may be helpful for better organizing, discovering and retrieving items [17, 21, 22]. The current work focuses on a relevant yet different application of CTSes, that is, to provide personalized item

**Zi-Ke Zhang, Tao Zhou and Yi-Cheng Zhang**

recommendations with the help of tag information. Schenkel et al. [23] proposed an incremental threshold algorithm taking into account both the social ties among users and semantic relatedness of different tags, which performs remarkably better than the algorithm without tag expansion. Nakamoto et al. [24] created a tag-based contextual collaborative filtering model, where the tag information is treated as the users' profiles. Tso-Sutter et al. [25] proposed a generic method that allows tags to be incorporated to the standard collaborative filtering, via reducing the ternary correlations to three binary correlations and then applying a fusion method to re-associate these correlations.

Some diffusion-based algorithms are recently proposed for personalized recommendations. Huang et al. [9] proposed a spreading activation based collaborative filtering, which is essentially an iterative diffusion process. This algorithm can provide relatively accurate recommendations for very sparse systems that are hard to be managed by the standard collaborative filtering. Zhou et al. [10] proposed an extremely fast algorithm considering only a two-step diffusion in user-item bipartite networks, which can still give slightly more accurate recommendations than the standard collaborative filtering. Further more, some refinement taking into account the initial difference of activations for different items can simultaneously enhance the recommendation accuracy and diversification [26]. Zhang et al. [27] proposed an iterative opinion diffusion algorithm to predict ratings in Netfix.com. All the above diffusion-based algorithms obey the conservation law, in contrast to which Zhang et al. [28] proposed a non-conservation diffusion-based recommendation algorithm mimicking the heat conduction in networks, which is very efficient to dig out the unpopular yet relevant items. Song et al. [29] proposed a so-called *DiffusionRank* algorithm where the prediction score is given by the likelihood that information can propagate from a given user to a given item within a certain time period. Liu et al. [30] proposed a diffusion-based top-k collaborative filtering, which performs better than pure top-k CF and pure diffusion-based algorithm.

The rest of this paper is organized as follows. In Section 2, we introduce the diffusion process on bipartite graphs, as well as our proposed integrated algorithm on tripartite graphs. Section 3 describes the data sets and reports the experimental results, including *the area under the receiver operating characteristic* (ROC) *curve* [31, 32], *Recall* [11], *Diversification* [26] and *novelty* [26]. Finally, we summarize this paper and outline some open issues for future research in Section 4.

## 2. Method

We start by introducing a diffusion-based recommendation algorithm on bipartite graphs, and then show the integrated algorithm on tripartite graphs. A simple example will be given to help readers well and truly understand the algorithmic procedure. A recommender system considered in this paper consists of three sets, respectively of users $U = \{U_1, U_2, \ldots, U_n\}$, items $I = \{I_1, I_2, \ldots, I_m\}$, and tags $T = \{T_1, T_2, \ldots, T_r\}$. The tripartite graph representation can be described by two adjacent matrices, $A$ and $A'$, for user-item and item-tag relations. If $U_i$ has collected $I_j$, we set $a_{ij} = 1$, otherwise $a_{ij} = 0$. Analogously, we set $a'_{jk} = 1$ if $I_j$ has been assigned by the tag $T_k$, and $a'_{jk} = 0$ otherwise. Figure 2 shows an illustration consisted of three users, five items and four tags.

### 2.1. Diffusion on Bipartite Graphs

Considering a bipartite graph $G(U, I, E)$, where $U$ and $I$ are user set and item set, and $E$ is the set of edges connecting users and items. Supposing that a kind of resource is initially located on items, each item will averagely distribute its resource to all neighboring users, and then each user will redistribute the received resource to all his/her collected items. Denoting $\vec{f}$ the initial resource vector on items (i.e., $f_j$ is the amount of resource located on $I_j$), then the final resource vector, $\vec{f'}$, after the two-step diffusion is:

**Zi-Ke Zhang, Tao Zhou and Yi-Cheng Zhang**

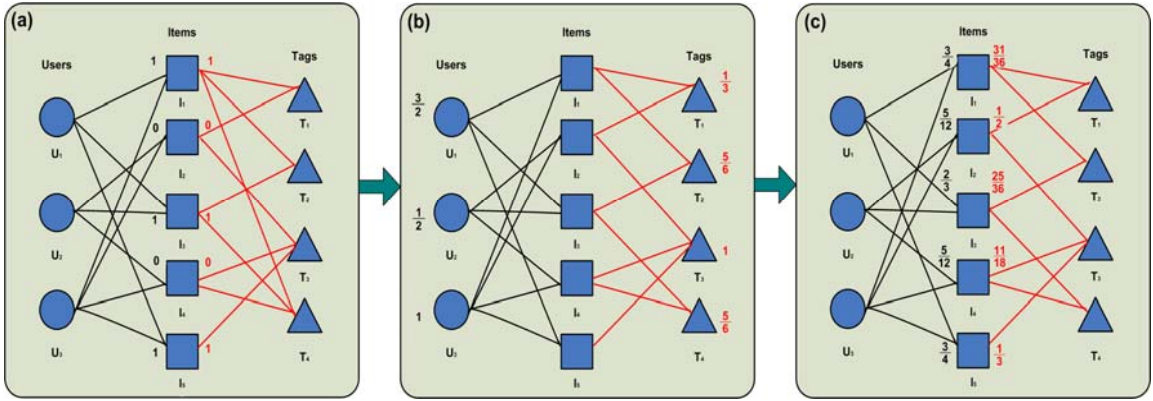

Fig.2. Illustration of the integrated diffusion process on a tripartite graph consisted of three users, five items and four tags. Plot (a) shows the initial condition given U1 as the target user, plot (b) describes the result after first-step diffusion, during which the resources are transferred from items to users and tags. Eventually, the resources flow back to items, and we show the result in plot (c).

$$f'_j = \sum_{l=1}^{n} \frac{a_{lj}}{k(U_l)} \sum_{s=1}^{m} \frac{f_s}{k(I_s)}, \quad j=1,2,\cdots,m, \qquad (1)$$

where $k(U_l) = \sum_{j=1}^{m} a_{lj}$ is the number of collected items for user $U_l$, and $k(I_s) = \sum_{i=1}^{n} a_{is}$ is the number of neighboring users for item $I_s$.

Given a target user $U_i$, we set the initial resource vector, $\vec{f}(U_i)$, as

$$f_j(U_i) = a_{ij}, \quad j=1,2,\cdots,m, \qquad (2)$$

In this case, the initial resource can be understood as giving a unit recommending capacity to each collected item, and the different initial resource vectors for different users have captured the personalized preferences. The final resource vector, $\vec{f'}(U_i)$, is obtained by Eq. (1). All $U_i$'s uncollected items are sorted in the descending order of final resource, and those items with highest values of resource are recommended. This algorithm is originally motivated by the resource-allocation process on graphs [33], and has been shown to be slightly more accurate than the collaborative filtering based on MovieLens data [10].

## 2.2. Integrated Diffusion on Tripartite Graphs

As mentioned in Section 2.1, the different collections of items reflect personalized preferences. Further more, even for exactly the same items, different users may assign far different tags to them. Therefore, the significance of collaborative tags in a recommender system is twofold. Firstly, tags richen the item information and two items sharing many common tags are probably close related in content. Secondly, the personalized preferences are naturally embedded in the different usages of tags. In this subsection, we introduce a simple way that utilizes the tag information to provide better recommendations.

We first consider the diffusion on an item-tag bipartite graph. Analogously, supposing that a kind of resource is initially located on items, each item will equally distribute its resource to all neighboring tags, and then each tag will redistribute the received resource to all its neighboring items. Thus, given the initial resource vector, $\vec{f}$, the final resource vector, $\vec{f''}$, is:

$$f''_j = \sum_{l=1}^{r} \frac{a'_{jl}}{k(T_l)} \sum_{s=1}^{m} \frac{f_s}{k'(I_s)}, \quad j=1,2,\cdots,m, \tag{3}$$

where $k(T_l) = \sum_{j=1}^{m} a'_{lj}$ is the number of neighboring items for tag $T_l$, $k'(I_s) = \sum_{i=1}^{r} a'_{sl}$ is the number of neighboring tags for item Is.

Given a target user $U_i$, the initial resource is also set according to Eq. (2). As a start point, in this paper, we adopt the simplest way to integrate the diffusions on user-item and item-tag bipartite graphs, that is, to define the final resource as a linear superposition of $\vec{f'}$ and $\vec{f''}$:

$$\vec{f^*} = \lambda \vec{f'} + (1-\lambda)\vec{f''}, \tag{4}$$

where $\lambda \in [0,1]$ is a tunable parameter, $\vec{f'}$ is the vector obtained from Eq. (1), $\vec{f''}$ is the vector derived from Eq.(3), and both of the two vectors are attained for the same target user. In the extremal cases $\lambda = 0$ and $\lambda = 1$, the integrated algorithm degenerates to the pure diffusions on item-tag and user-item bipartite graphs, respectively.

## 2.3. Example

In this subsection, we give a simple example to concretely describe the above integrated diffusion process. As shown in Figure 2, $U_1$ is chosen to be the target user. Firstly, we highlight all the items collected by $U_1$ and assign each of them a unit resource. Thus the initial vector, as shown in Figure 2(a), is:

$$\vec{f} = (1,0,1,0,1). \tag{5}$$

Secondly, the initial resource will be transferred to users and tags (treated as two independent processes), resulting in two medi-distributions of resources on users and tags, as shown in Figure 2(b). Finally, the resources located on users and tags will be transferred back to items. As shown in Figure 2(c), the final resource vectors, corresponding to Eq. (1) and Eq. (3), are:

$$\vec{f'} = (\frac{3}{4},\frac{5}{12},\frac{2}{3},\frac{5}{12},\frac{3}{4}),$$
$$\vec{f''} = (\frac{31}{36},\frac{1}{2},\frac{25}{36},\frac{11}{36},\frac{1}{3}). \tag{6}$$

Given ,, the recommendation scores on items of $U_1$ can be obtained by Eq. (4). The items with highest scores are then recommended to $U_1$.

**Zi-Ke Zhang, Tao Zhou and Yi-Cheng Zhang**

Table 1.

Basic statistics of three data sets. $n$, $m$, $r$ are the total numbers of users, items and tags, respectively. $<k>$ denotes the average number of users having collected an item, and $<k'>$ refers to the average number of tags assigned to an item.

| Data set | $n$ | $m$ | $r$ | $<k>$ | $<k'>$ |
|---|---|---|---|---|---|
| *Del.icio.us* | 11999 | 285804 | 119392 | 5.58 | 9.46 |
| *MovieLens* | 3549 | 6054 | 5828 | 8.45 | 10.00 |
| *BibSonomy* | 794 | 11211 | 5986 | 2.74 | 7.91 |

## 3. Experiment

### 3.1. Data Sets

In this paper, three representative data sets, Del.icio.us[1], MovieLens[2] and BibSonomy[3], are used to evaluate our proposed algorithm. *Del.icio.us* is one of the most popular social bookmarking web sites, which allows users not only to store and organize personal bookmarks (URLs), but also to look into other users' collections and find what they might be interested in by simply keeping track of the baskets with tags or items. The data used in this paper is a random sampling of about 12000 uses in May 2008. *MovieLens* is a movie rating system, where each user votes movies in five discrete ratings 1-5. A tagging function is added in from January 2006. *BibSonomy* is a collaborative tagging system mixing with social bookmarks and scientific publications [34], and we only choose the data of bookmarks. In all the three data sets, we remove the isolated nodes and guarantee that each user has collected at least one item, each item has been collected by at least two users, and assigned by at least one tag. *MovieLens* and *BibSonomy* allow users freely to assign tags without any grammatical restrictions, resulting in many strange and/or extremely long tags, some of which are phrases, and others may contain rarely used symbols. Therefore, for *MovieLens* and *BibSonomy*, the tags appearing only once are removed. Table 1 summarizes the basic statistics of the three purified data sets.

### 3.2. Metrics for Algorithmic Performance

Every data set is consisted of many entries, each of which follows the form {user, item, $tag_1$, $tag_2$, ⋯ , $tag_h$}, where $h$ is the number of tags assigned to the relevant item by the very user. To test the algorithmic performance, each data set is randomly divided into two parts: the training set contains 95% of entries, and the remaining 5% of entries constitutes the testing set. The training set is treated as known information used for generating recommendations, while no information in the testing set is allowed to be used for recommending. To give solid and comprehensive evaluation of the proposed algorithm, we employ four different metrics that characterizing not only the accuracy of recommendations, but also the diversification and novelty.

---

[1] http://del.icio.us
[2] http://movielens.org
[3] http://www.bibsonomy.org

Table 2.

Comparison of algorithmic accuracy, measured by the AUC. *Pure U-I* and *Pure I-T* denote the pure diffusions on user-item bipartite graphs and item-tag bipartite graphs, respectively corresponding to $\lambda=1$ and $\lambda=0$. The optimal values of $\lambda$ as well as the corresponding optima of AUC are presented.

| Data set | Pure U-I | Pure I-T | Optimum | $\lambda_{opt}$ |
|---|---|---|---|---|
| *Del.icio.us* | 0.8098 | 0.8486 | 0.8588 | 0.32 |
| *MovieLens* | 0.8065 | 0.8163 | 0.8233 | 0.44 |
| *BibSonomy* | 0.7374 | 0.7600 | 0.7852 | 0.44 |

- The area under the ROC curve [31, 32]: In the present case, the area under the ROC curve, abbreviated by AUC, for a particular user is the probability that a randomly selected removed item for this user (i.e., an item in the testing set and being collected by this user) is given a higher score by our algorithm than a randomly selected uncollected item (i.e., an item irrelevant to this user in neither the training set nor the testing set). The AUC for the whole system is the average over all users. If all the scores are generated from an independent and identical distribution, AUC $\approx$ 0.5. Therefore, the degree to which the AUC exceeds 0.5 indicates how much better the algorithm performs than pure chance.
- Recall [11]: Note that, the AUC takes into account the order of all uncollected items, however, in the real applications, user might only care about the recommended items, that is, the items with highest scores. Therefore, as a complementary measure, recall is employed to quantify the accuracy of recommended items, which is defined as:

$$Recall = \frac{1}{n}\sum_{i=1}^{n} N_r^i / N_p^i, \quad 0 \leq Recall \leq 1, \tag{7}$$

where $N_p^i$ is the number of items collected by $U_i$ in the testing set, and $N_r^i$ is the number of recovered items in the recommendations for $U_i$. We use the averaged recall instead of simply counting $N_r / N_p$ with $N_r = \sum_i N_r^i$ and $N_p = \sum_i N_p^i$ since it is fair to give the same weight on every user in the algorithm evaluation. Assuming the length of recommendation list, $L$, is fixed for every user, recall is very sensitive to L and a larger $L$ generally gives a higher recall.
- Diversification [26]: This measure considers the uniqueness of different users' recommendation lists, thus can be understood as the inter-user diversity[4]. Denote $I_R^i$ the set of recommended items for user $U_i$, then

---

[4] To evaluate the diversification of recommendations, Ziegler *et al.* [35] proposed a metric for intra-list diversity, which requires a similarity measure for item pairs in the recommendation list. This similarity can be well defined with the help of item attributes or item contents. However, that information is not available in general. Of course, one can define a kind of similarity between two items by counting the number of users collected both of them and/or the number of tags shared by both of them. However, how to weight the contributions to similarity from the user-item relations and item-tag relations is a problem, and the similarity measure itself has latent correlation with the algorithm (for they both have used the same information) that may lead to systematic bias. We therefore employ the inter-user diversification, which is simple and meaningful for any kind of data.

$$Diversification = \frac{2}{n(n-1)}\sum_{i \neq j}(1-\frac{|I_R^i \cap I_R^j|}{L}), \qquad (8)$$

where $L = |I_R^i|$ for any $i$ is the length of recommendation list. Greater or lesser values of diversification mean respectively greater or lesser personalization of users' recommendation lists.

- Novelty [26]: This measure quantifies the capacity of an algorithm to generate novel and unexpected results, that is to say, to recommend less popular items unlikely to be already known about. The simplest way is to use the average collected times over all recommended items, as:

$$Novelty = \frac{1}{nL}\sum_{i=1}^{n}\sum_{I_r \in I_R^i} k(I_r), \qquad (9)$$

with $k(I_r)$ the number of users having collected the item $I_r$. See also previous definition in Eq. (1). Readers are warned that the less value obtained by Eq. (9) actually means higher novelty and surprisal[5].

### 3.3. Experimental Results

Figure 3 shows the AUC statistics of the three data sets for $\lambda \in [0,1]$. It can be seen that all three curves have qualitatively the same trend: (i) pure diffusion on item-tag bipartite graphs ($\lambda = 0$) performs better that on user-item bipartite graphs ($\lambda = 1$), (ii) integrated algorithm can pro- vide more accurate recommendations than the pure cases. In Table 2, we compare the AUC for pure diffusions on bipartite graphs and for the optimum by integrated algorithm. Comparing with the algorithm without tag information [10], at the optimal values, the improvements for *Del.icio.us, MovieLens* and *BibSonomy* are 6.1%, 2.1% and 6.5%, respectively. Since AUC is not a very sensitive index [32], 6% is indeed a remarkable improvement.

In Figure 4, we report the values of recall. Since the typical length for recommendation list is tens, our experimental study focuses on the interval $L \in [10,100]$. To keep the figure neat, we only show two pure diffusions with $\lambda = 0$ and $\lambda = 1$, as well as a naive integration with $\lambda = 0.5$. Different from the AUC statistics, the diffusion on item-tag bipartite graphs does not always perform better than the one on user-item bipartite graphs. However, the integrated algorithm can always beat the two pure algorithms in certain ranges of ߣ, which again indicates the advantage of integration.

Figure 5 and Figure 6 report the diversification and novelty of recommendations versus the parameter ߣ, respectively. All curves are monotone, and the algorithm depending more on tags can provide more personalized and higher novel recommendations. That is to say, the collaborative tagging system can be used to simultaneously enhance the serendipitous discovering of individuals and the diversity among individuals.

### 4. Conclusion and Discussion

In this paper, we proposed an integrated diffusion-based algorithm with the help of collaborative tagging information. Experimental results demonstrate that the usage of tag information can significantly improve accuracy, diversification and novelty of recommendations. It is worthwhile to emphasize the twofold significance of the collaborative tags, as an assistant part of a recommender system. Firstly, tags assigned to a certain item can be considered as the highly abstracted content of this item. Secondly, even for the same item, different user may assign completely different tags, and thus the personalized preferences are naturally embedded in the different usages of tags. A user assigns two tags, action and Nicolas Cage, to the movie Lord of War may get a recommendation Gone in Sixty Seconds, while another user sticks drama film and peace onto the same movie may be recommended Crash.

---

[5] An alternative is to user $k^{-1}(I_r)$ instead of $k(I_r)$ in Eq. (9), then the larger value corresponds to higher novelty. We employ the current definition for the advantage to directly exhibit the average collected times of recommended items.

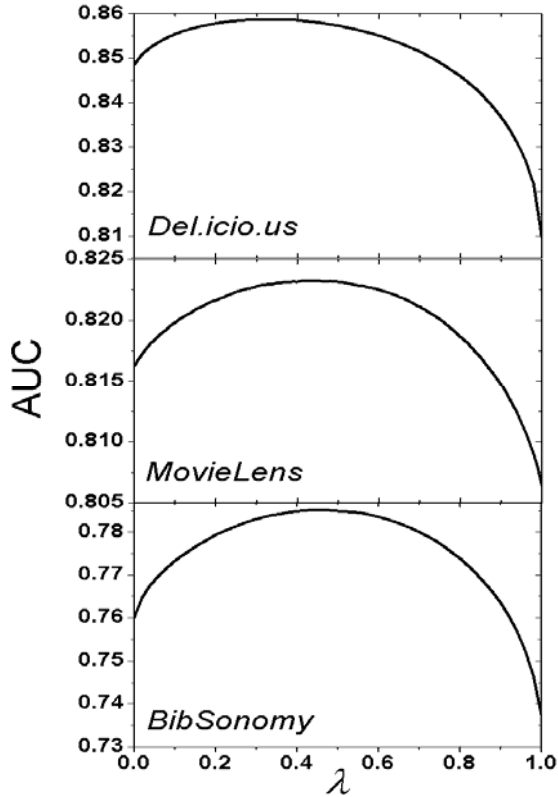

Fig.3. *AUC* versus $\lambda$. The results reported here are averaged over 50 independent runs, each of which corresponds to a random division of training set and testing set.

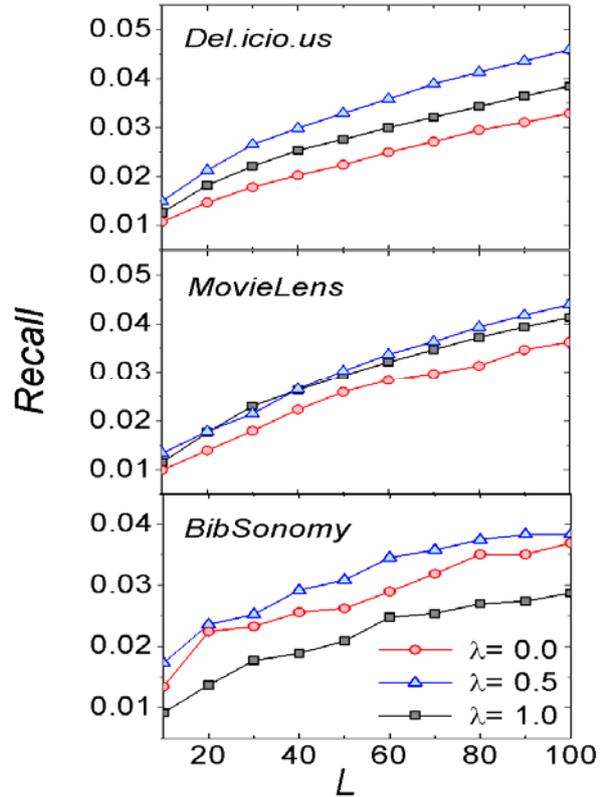

Fig.4. Recall versus the length of recommendation list for $\lambda = 0$, $\lambda = 0.5$ and $\lambda = 1.0$.

The collaborative tagging systems play more and more important role in the Internet world, and we must be aware of their significance. Experimental results in this paper strongly suggest using the tag information to improve the quality of recommendations. Indeed, we should do even beyond, that is, for the already existed recommender systems, to add tagging functions into them and encourage users to organize their collections by using tags.

This paper only provides a simple start point for the design of hybrid algorithms making use of tag information, and a couple of open issues remain for future study. First, we lack quantitative understanding of the structure and evolution of collaborative tagging systems as well as the performance of folksonomy. Although the relation between folksonomy and recommender systems is not clear thus far, we deem that the in-depth understanding of tagging systems should be helpful for better recommending. Second, the current algorithm focus

**Zi-Ke Zhang, Tao Zhou and Yi-Cheng Zhang**

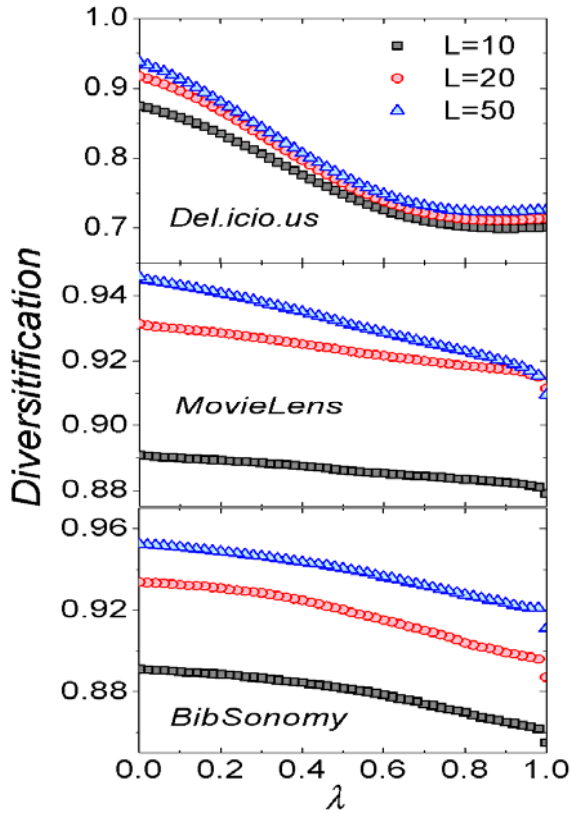

Fig.5. *Diversification* versus $\lambda$ for typical recommendation list lengths: $L = 10$, $L = 20$ and $L = 50$.

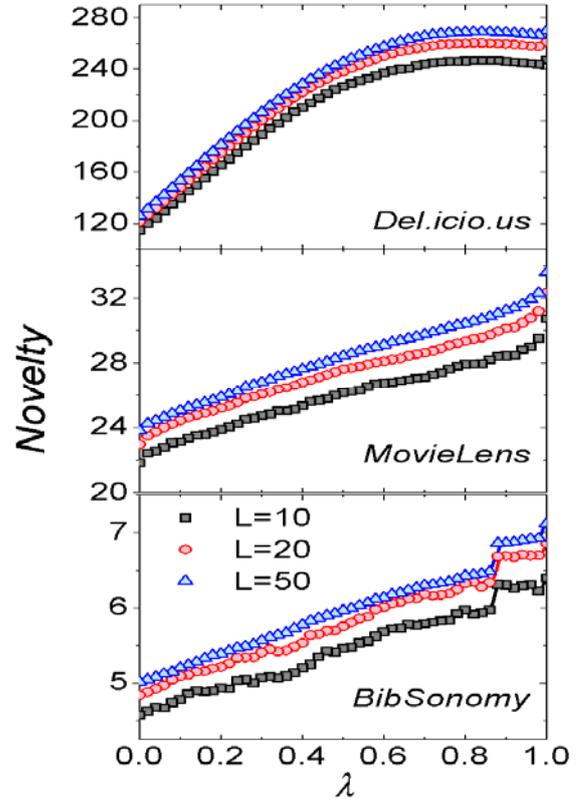

Fig.6. *Novelty* versus $\lambda$ for typical recommendation list lengths: $L = 10$, $L = 20$ and $L = 50$.

on the unweighted graphs, and a refined algorithm with properly defined weights on user-item and item-tag relations may further improve the algorithmic performance. Third, it is significant to design an on-line algorithm that can give real-time response to the user activities, such as selecting of new items and changing of tags. Finally, the tag information can also be exploited under the frameworks of collaborative filtering [11] and iterative diffusion algorithm [9], as well as some more complicated methods such as Probabilistic Latent Semantic Analysis [36], Latent Dirichlet Allocation [37] and Iterative Latent Semantic Analysis [38]. Systematic investigation on tag-aware recommendation algorithms must be very helpful in the futuredesign of recommender systems.

## 5. Acknowledgement

We acknowledge the GroupLens Research Group for MovieLens data, and the Knowledge and Data Swiss National Science Foundation (Project 205120-113842 and 200020-121848). T.Z. acknowledges the National Natural Science Foundation of China under Grant No. 60744003.

**Zi-Ke Zhang, Tao Zhou and Yi-Cheng Zhang**